\documentclass{aastex631}

\usepackage{upgreek}

\begin{document}

\title{Merger-Driven Turbulence and Coherent Transport in the Intracluster Medium 
}

\author[0009-0005-2983-2729]{Yiwei. Zhang}
\affiliation{South-Western Institute for Astronomy Research (SWIFAR) \\ Yunnan University, 650500 Kunming, P. R. China; zyw11349@gmail.com}

\author[0000-0003-2076-4510]{Xun Shi}
\affiliation{South-Western Institute for Astronomy Research (SWIFAR) \\ Yunnan University, 650500 Kunming, P. R. China; xun@ynu.edu.cn}

\author[0000-0002-6766-5942]{Daisuke Nagai}
\affiliation{Department of Physics, Yale University\\  New Haven, CT 06520, USA; daisuke.nagai@yale.edu}



\begin{abstract}
The distribution of metals and temperature in the intracluster medium (ICM) provides key insights into galaxy cluster evolution, revealing information about chemical enrichment and heating and cooling processes, respectively. To access this information, it is crucial to understand the transport processes in the ICM. Here, we systematically study the transport mechanisms in the ICM with tracer particle resimulations of the Omega500 cosmological hydrosimulation, using a sample of four galaxy clusters of comparable masses but different mass assembly histories. Through the analysis of particle pair dispersion statistics, we find a time-dependent scaling index linked to the cluster's dynamical state. It reaches or exceeds Richardson scaling briefly during major mergers but remains much lower in relaxed clusters. We identify a coherent transport mode during major mergers that causes directional flow in the ICM. Although coherent transport can move particles to outer regions, the particles transported to the cluster outskirts compose only a small fraction of the density there; thus the anisotropy it creates in the overall density distribution is limited. Moreover, strong turbulence generated by mergers quickly disperses these particles, further limiting this effect. We also provide useful statistics on the radial evolution of the ICM and the fraction of particles that ever reached the inner regions as a function of radius. Our results show that major mergers primarily drive particle transport, linking ICM transport to merger-driven dynamics, and highlighting the interplay between coherent and turbulent transport.

\end{abstract}



\section{Introduction}\label{sec:intro}

Galaxy clusters are the largest gravitationally bound systems in the Universe primarily composed of dark matter and the intracluster medium (ICM), and with a small mass contribution from the galaxies. The formation of galaxy clusters is one of the most intense astrophysical processes, involving gravitational collapse and a variety of complex physical mechanisms associated with galaxy formation. Therefore, galaxy clusters hold an indispensable position in cosmological and astrophysical research. By studying galaxy clusters in depth, we can uncover the evolutionary patterns of large-scale cosmic structures, analyze the dynamics of the ICM, and deepen our understanding of plasma physics and the processes of galaxy formation. Additionally, the observation and simulation of galaxy clusters provide crucial constraints on key cosmological parameters. The ICM is a hot, ionized gas predominantly made up of hydrogen and helium with trace amounts of heavier elements. It exhibits extremely high temperatures, typically ranging from $10^{7}$ to $10^{8}$ K, and emits significantly in the X-ray spectrum. The dynamic and thermodynamic properties of the ICM play a crucial role in the overall structure and evolution of galaxy clusters.

During the evolution of galaxy clusters, their structures are not static but change over time due to gravitational interactions. The ICM is characterized by turbulent, random motions and bulk flows resulting from cluster mergers and interactions (\citealt{Lau2009, Vazza2011, schmidt2017, Ha2018, Li2018, vazza2021}). Galaxy cluster mergers are among the most violent events in cluster evolution and are crucial mechanisms impacting ICM dynamics. The disturbance of gas during mergers is a key driver of turbulence in the ICM \citep{ Norman1998, Iapichino2008, Nelson2012, Vazza2012, Nagai2013, Miniati2014, Iapichino2017, Valdarnini2011, Valdarnini2019, Vazza2017}. 

The presence of turbulence in the ICM has been observationally confirmed through various methods. The Hitomi satellite provided direct evidence of turbulence by detecting the non-thermal broadening of X-ray emission lines (\citealt{Hitomi2016}). Additionally, several indirect methods have demonstrated the existence of turbulence, such as X-ray surface brightness fluctuations or pressure fluctuations inferred from X-ray maps and Sunyaev-Zel’dovich effect maps (\citealt{Schuecker2004, churazov2012, Walker2015, Khatri2016, zhuravleva2018}), magnetic field fluctuations in diffuse cluster radio sources (\citealt{Murgia2004, Vogt2005, Bonafede2010, Vacca2010, Vacca2012}), and suppression of resonant line scattering in X-ray spectra (\citealt{Churazov2004, Zhuravleva2013, Hitomi2018}). Turbulent gas motions also contribute to non-thermal pressure \citep{Lau2009, Battaglia2012, Nelson2014, Shi2014, shi2015}, leading to deviations in galaxy cluster mass estimates that assume hydrostatic equilibrium \citep{Rasia2006, Nagai2007, Piffaretti2008, Battaglia2012, Lau2013, Biffi2016, shi2015, Henson2016, Ansarifard2020}. With increasingly high-resolution satellites, such as XRISM and Athena, more direct measurements of ICM turbulence will become feasible.

Turbulence can lead to super-diffusion whose efficiency is significantly higher than that of thermal diffusion (e.g., \citealt{Ragot1997, Shtykovskiy_2010, Perri2012, Zimbardo2013, Bykov2017, Effenberger2024}). Therefore, studying turbulent transport is critical to a better understanding of the distribution and mixing of metals in the ICM (\citealt{Vazza_2010, Ruszkowski2010, Werner2010, Ezer2017, Urban_2017, Sarkar_2022}). Moreover, heat transport induced by turbulent mixing can counteract the radiative cooling of the hot gas, making it a potentially crucial factor in addressing the problem of gas cooling in cluster cores (\citealt{Kim2003, Zhuravleva2014}). 

The transport in the ICM may include not only turbulence transport. During major merger events, a coherent transport mechanism may exist. If such a coherent transport mode exists, it could collectively move particles from the inner regions of the galaxy cluster to its outskirts along a specific direction, and it could play a role in explaining the observed metal abundance in the ICM periphery within the galaxy cluster (\citealt{Fujita2008PASJ...60S.343F, Urban_2017, Ezer2017}). In this work, we systematically study the transport mechanisms in ICM using numerical simulations. We focus on the transport properties in the ICM during the evolution of galaxy clusters, particularly during major merger events, and study the role of turbulence on transport in the ICM.

\section{Turbulent Particle Dispersion in Galaxy Clusters}\label{sec: tho}
Particle pair dispersion is a fundamental characteristic of turbulent transport, describing the rate at which two particles separate over time. It also serves as a tool for quantifying the variance of the concentration fluctuations of substances carried by the fluid. In a quiescent fluid, the relative dispersion between particles is dominated by diffusion. The particles follow a Brownian motion, with their mean-square separation growing linearly with time. However, in turbulence, the dispersion is driven by advection, fundamentally altering the nature of the process.

The evolution of particle pair dispersion in turbulence can be divided into three stages, which have been extensively studied (\citealt{Monin1977, Sawford2001, Salazar2009, Shnapp2023}). The first stage, the ``ballistic stage'', occurs over very short timescales. During this stage, the relative velocity between the particles remains approximately constant as a result of the finite inertia of the fluid, leading to a linear increase in their separation distance with time. The duration of this stage is determined by the Batchelor timescale \( \tau_b \), defined as:

\begin{equation}
\tau_b = \left(\frac{l_0^2}{\epsilon}\right)^{1/3}
\end{equation}
where \( l_0 \) is the initial separation between particles, and \( \epsilon \) is the average energy dissipation rate.

The second stage occurs within the inertial range of turbulence, characterized by \( \eta \ll l \ll L \), where \( L \) is the integral length scale, and \( \eta \) is the Kolmogorov length scale, defined as:

\begin{equation}
\eta = \left(\frac{\nu^3}{\epsilon}\right)^{1/4}
\end{equation}
where \( \nu \) represents the kinematic viscosity. During this ``inertial stage'', the relative velocity between particles increases with their separation distance, resulting in super-diffusive growth. Richardson (\citealt{Richardson1926}) was the first to propose a scaling law for this stage. 
Based on Kolmogorov’s turbulence theory, Obukhov clarified that the pair dispersion in the inertial range grows as \( g \epsilon t^3 \)\citep{Obukhov1959}. Here, \( g \) is a universal constant, and the only relevant turbulent parameter is the energy dissipation rate per unit mass \( \epsilon \).

As the separation distance \( l \) exceeds the inertial range of the flow (\( l \gg L \)), the third stage, the ``diffusion stage'', begins. In this stage, the influence of the turbulent velocity field becomes negligible, and the particle dispersion resembles Brownian motion, with the separation distance growing proportionally to the square root of time.

This classical picture based on homogeneous, steady-state turbulence can guide our study of turbulent transport within galaxy clusters. However, galaxy clusters exhibit highly dynamic environments that are continuously influenced by mergers and accretion, and the ICM has density stratification that affects the properties of its turbulence\citep{Shi_2019, Mohapatra2021, Wang2023}. \citet{Vazza_2010} statistically analyzed the pair dispersion in the hot ICM gas and found that the particles exhibit super-diffusive behavior that largely follows the Richardson scaling \( <\Delta x^2> \propto t^3 \). However, since turbulence in galaxy clusters is neither homogeneous nor steady-state, interpreting such a finding is not straightforward.

It is also important to note that the hydrodynamic Reynolds number of the ICM estimated using the classical Braginskii viscosity (i.e., determined by thermal ion-ion Coulomb collisions) is not high \citep{Brunetti2014}. Magnetohydrodynamic turbulence is generally expected to be generated due to the relatively large magnetic Reynolds number (\citealt{Murgia2004, Vogt2005, Bonafede2010, Vacca2010, Vacca2012}). Additionally, numerical simulations are limited by resolution constraints, leading to relatively low effective Reynolds numbers and small inertial ranges. Given the theoretical uncertainty and numerical constraints, our study focuses on the effects of fluid motions on diffusion at relatively large scales.

\section{Simulation} \label{sec:style}

Our research is based on the non-radiative run of the Omega500 simulation \citep{Nelson_2014}, a cosmological Eulerian simulation specifically designed to study the evolution of galaxy clusters. The Omega500 simulation serves as a powerful tool for understanding the dynamics of the ICM within galaxy clusters. The simulation employed the Adaptive Refinement Tree (ART) N-body + hydrodynamics code (\citealt{Kravtsov1999, Kravtsov2002, Rudd2008}). This Eulerian-based code incorporates adaptive refinement in both spatial and temporal dimensions while maintaining a fixed mass resolution \citep{Klypin2001}. This approach ensures the dynamic range required to resolve halo cores in self-consistent cosmological simulations.

The simulation adopted a flat $\Lambda$CDM cosmology based on the five-year WMAP (WMAP5) cosmological parameters: $\Omega_{\rm{m}} = 0.27$, $\Omega_{\rm{b}} = 0.0469$, $h = 0.7$, and $\sigma_8 = 0.8$, where $h$ is the normalized Hubble constant, and $\sigma_8$ represents the mass variance within spheres of radius 8$h^{-1}$ Mpc. The simulation was conducted within a comoving box of size $500 h^{-1} \, \text{Mpc}$, using a uniform $512^3$ grid with eight levels of refinement, achieving a maximum comoving spatial resolution of $3.8 h^{-1} \, \text{kpc}$.

Galaxy clusters with masses $M_{500\rm{c}} \geq 3 \times 10^{14} h^{-1} M_\odot$ were selected, and high-resolution simulations were performed in the regions surrounding these clusters. The final effective resolution of the simulation reached $2048^3$, achieving a mass resolution of $1.09 \times 10^9 h^{-1} M_\odot$ for the selected clusters. The non-radiative run of the simulation only included gravitational physics and non-radiative hydrodynamics. However, as \citet{Lau2009} pointed out, excluding cooling and star formation processes has minimal impact on the contribution of gas motions to the hydrostatic mass bias of galaxy clusters (less than a few percent), especially in regions outside the cluster cores. Magnetic fields, although known to impact microscopic transport (e.g., thermal conduction) in the ICM \cite[e.g.,][]{Komarov2016}, are weak and passive on large scales. On the 100 kpc - 10 Mpc scales that we consider in this study, the effects of radiative cooling or magnetic fields are expected to be negligible.

The Lagrangian method offers a significant advantage in tracking particle trajectories. This feature makes it particularly well-suited for studying particle transport mechanisms in the ICM of galaxy clusters. We utilized a Lagrangian tracer particle re-simulation of the Omega500 cosmological simulation (see \citealt{Shi_2020}). Lagrangian tracer particles were introduced into the Eulerian simulation at the initial conditions, with the number density of tracers proportional to the local gas density in regions surrounding galaxy clusters. These tracer particles were then passively carried along by the 3D velocity field of the Eulerian simulation as time progressed. During the output of tracer particle information, the thermodynamic properties of the highest resolution gas cell to which each particle belonged were recorded. The tracer particle simulation output was saved at a fine time resolution of 20-30 million years (Myrs). This approach allows for a detailed examination of particle transport processes in the ICM and sheds light on the dynamic behavior of the ICM within galaxy clusters.

\begin{figure*}
    \centering
    \begin{minipage}[b]{1\textwidth}
        \centering
        \includegraphics[width=\linewidth]{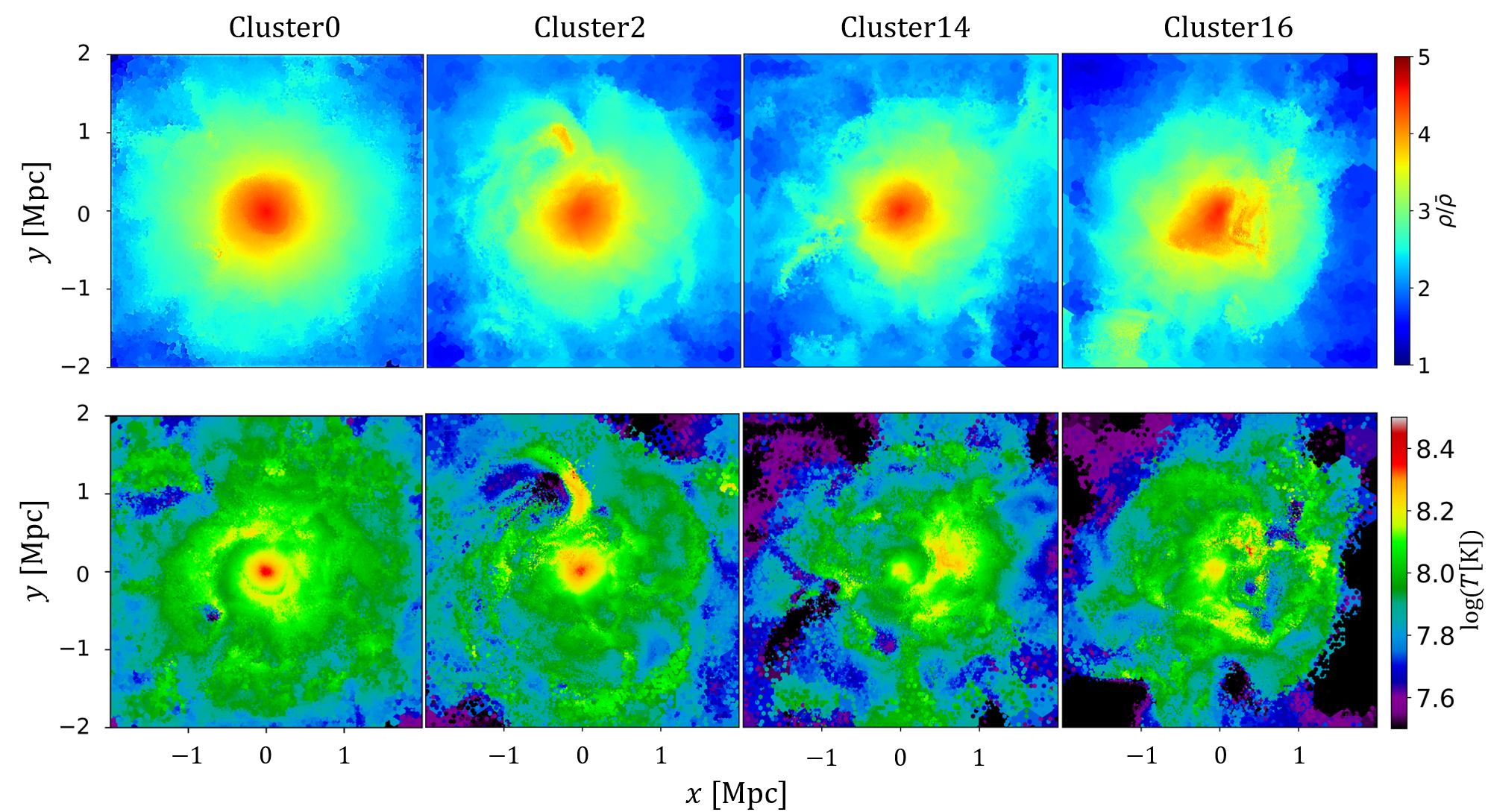}
    \end{minipage}
    \hspace{0.01\textwidth}
   
    \caption{The density (top) and temperature (bottom) distribution of the selected galaxy clusters at redshift $z=0$. Shown are slices of thickness 200 kpc across the center of the cluster. }
    \label{fig:rho}
\end{figure*}

\begin{figure}
    \includegraphics[width=\columnwidth]{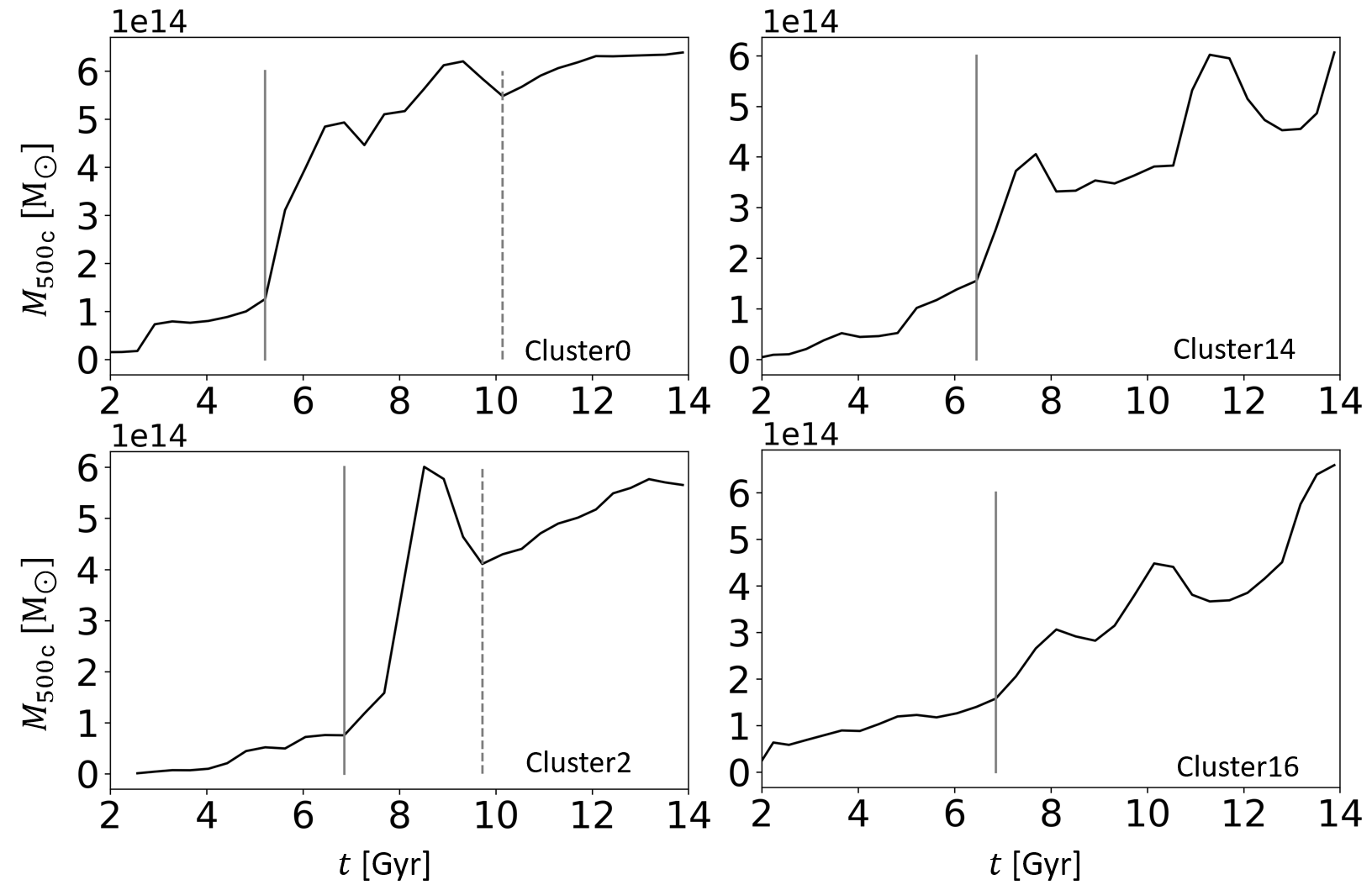}
    \caption{Mass growth histories of selected galaxy clusters. Each panel presents the evolution of $M_{500\rm{c}}$ as a function of time (t in Gyr) for four clusters with similar masses, labeled as Cluster0 (top left), Cluster14 (top right), Cluster2 (bottom left), and Cluster16 (bottom right). The solid lines indicate the onset of major merger events, and the dashed lines mark the end of the major merger phases. Despite their similar masses, each cluster exhibits distinct dynamical evolution characteristics. This variability in evolutionary paths enhances the robustness and general applicability of the results.}
    \label{fig:M-a}
\end{figure}

We selected four galaxy clusters as our study sample. Their density and temperature distributions at redshift $z=0$ are shown in Figure~\ref{fig:rho}. These four clusters were selected to have similar masses at $z=0$ but very different dynamical states and mass growth histories (see Figure~\ref{fig:M-a}). For Cluster0 and Cluster2, the major mergers had largely concluded by \( z=0 \), leaving the clusters in relatively relaxed states. For Cluster14 and in particular Cluster16, the major mergers were still ongoing at \( z=0 \), and these clusters remained dynamically unrelated. If observed in X-rays, they would likely appear as dynamically disturbed systems. In particular, they would exhibit lower core concentrations due to the redistribution of the ICM during mergers, and the w-parameter—reflecting the centroid shift—would be elevated, indicating significant dynamical activity. The basic information of these four clusters is summarized in Table\;\ref{tab:table}. In our work, we define the beginning of a major merger as the moment when the $R_{500\rm c}$ regions of the merging structures first come into contact. Correspondingly, $M_{500\rm{c}}$ experiences a sudden increase at this time. In our analysis, we studied the transport processes in these clusters in a region out to $10 R_{\rm 500c}$ to include the outskirts of the galaxy clusters. The total number of tracer particles in this volume was on the order of $10^6$.

\begin{deluxetable*}{lcccccc}
\tablecaption{Sample of simulated galaxy clusters \label{tab:table}}
\tablehead{
\colhead{Cluster} & 
\colhead{$R_{500\rm c}$ (Mpc)} & 
\colhead{$M_{500\rm c}$ ($M_\odot$)} & 
\colhead{Merger Start (Gyr)} & 
\colhead{Mass at Start ($M_\odot$)} & 
\colhead{Merger End (Gyr)} & 
\colhead{Mass at End ($M_\odot$)}
}
\startdata
Cluster0  & 1.31 & $6.39 \times 10^{14}$ & 5.22 & $1.26 \times 10^{14}$ & 10.14 & $5.48 \times 10^{14}$ \\
Cluster2  & 1.26 & $5.65 \times 10^{14}$ & 6.86 & $0.75 \times 10^{14}$ & 9.72 & $4.11 \times 10^{14}$ \\
Cluster14 & 1.29 & $6.06 \times 10^{14}$ & 6.46 & $1.55 \times 10^{14}$ & / & / \\
Cluster16 & 1.32 & $6.59 \times 10^{14}$ & 6.86 & $1.57 \times 10^{14}$ & / & / \\
\enddata
\end{deluxetable*}

\section{Results} \label{sec:floats}

\subsection{Pair dispersion statistics} \label{subsec:tables}
\begin{figure}
    \includegraphics[width=\columnwidth]{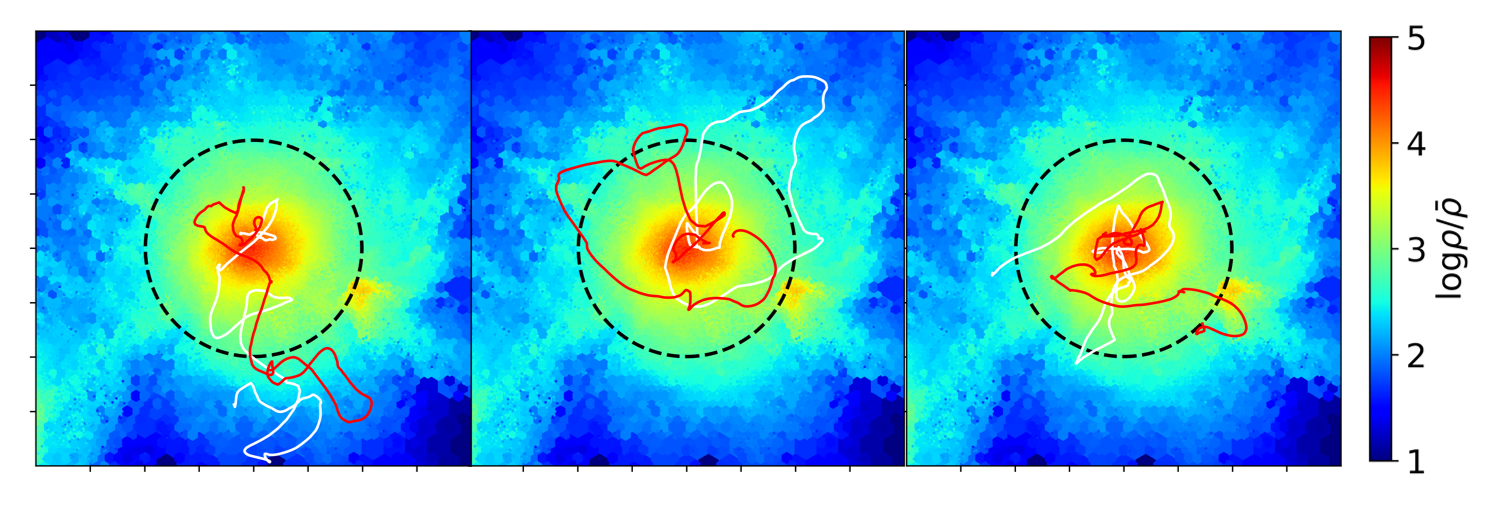}
    \caption{Examples of trajectories of particle pairs for Cluster0 during the period of $a=0.4-1$. The red and white lines distinguish between particles within the particle pairs, while the black circle denotes $R_{500\rm{c}}=0.4$Mpc at $a=0.4$.}
    \label{fig:trajec}
\end{figure}

To obtain the dynamical properties of galaxy clusters in our simulations, we studied the pair dispersion statistics of the tracer particles across cosmic time. Due to the turbulent motions in the ICM, pairs of particles originally close to each other will tend to separate with time. Figure~\ref{fig:trajec} illustrated this process by showing the projected trajectories of a few particle pairs. The speed with which the pairs of particles separated from each other depended statistically on the dynamical properties of the underlying medium. As mentioned in Section~\ref{sec: tho}, in a static thermalized medium, the particles exhibited a Brownian motion and the second moment of pairwise separation between the particles obeyed $<\Delta x^2> \propto t$ as a solution of the diffusion equation. However, in a medium with Kolmogorov turbulence, the particles exhibited super-diffusion in the inertial range under Richardson scaling $<\Delta x^2> \propto t^3$ \citep{Richardson1926}.

\citet{Vazza_2010} pioneered analysis of pair dispersion statistics of the ICM. They found that the pair dispersion increased following a behavior consistent with the Richardson scaling as soon as the cluster formation started, and they claimed this to be a general behavior found in all their clusters.

\begin{figure*}
    \centering
    \begin{minipage}[b]{1\textwidth}
        \centering
        \includegraphics[width=\linewidth]{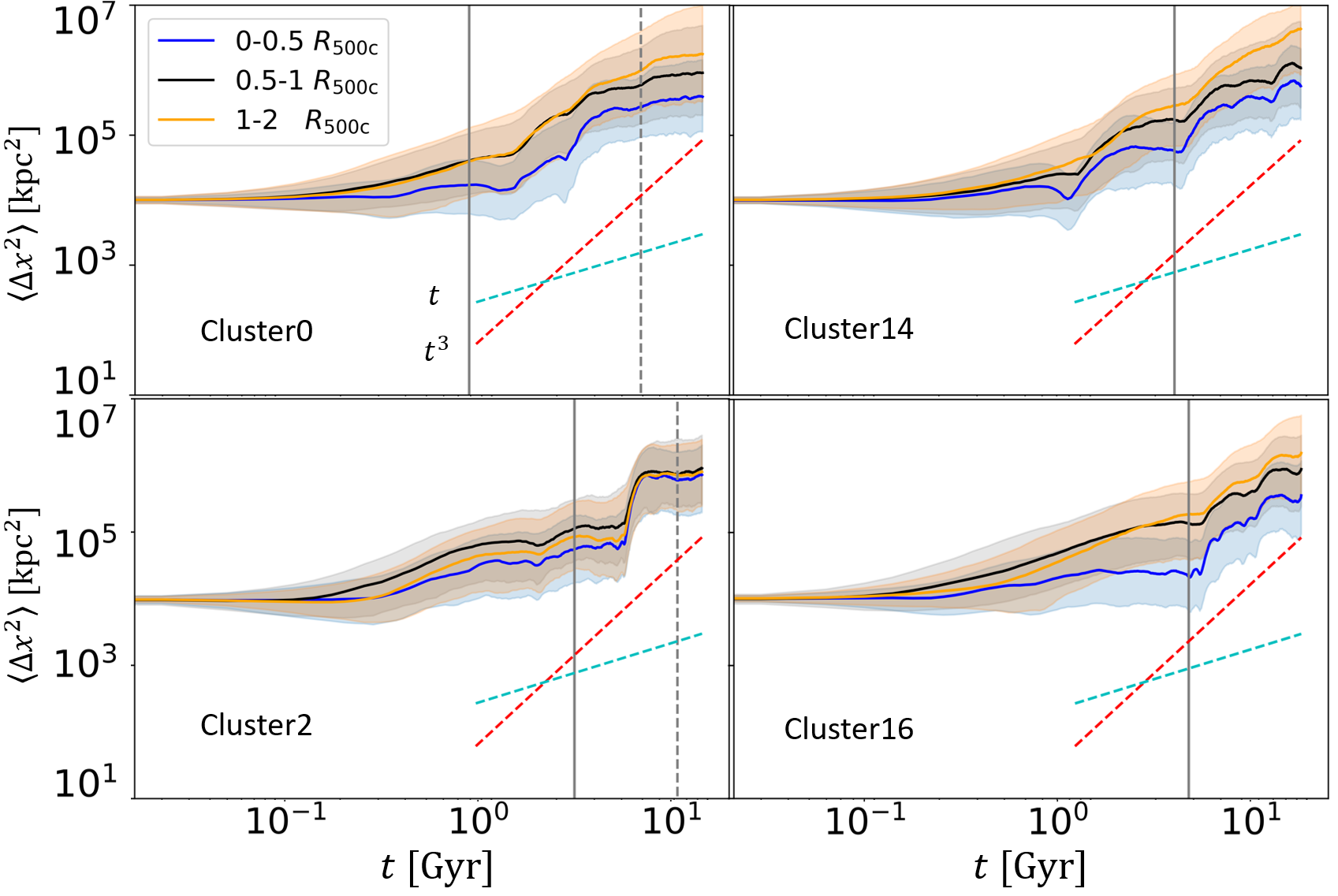}
    \end{minipage}
    \hspace{0.01\textwidth}
    \caption{Pair dispersion of particles of 4 simulated clusters starting at $a=0.4$.  Particles are grouped into three radial bins: 0-0.5$R_{500\rm{c}}$, 0.5-1$R_{500\rm{c}}$ and 1-2$R_{500\rm{c}}$, where $R_{500\rm{c}}$  is defined at $a=0.4$. The top row shows results for Cluster0 (left) and Cluster14 (right), while the bottom row presents results for Cluster2 (left) and Cluster16 (right). The red dashed line represents the Richardson scaling, $<\Delta x^2> \sim t^{3}$, and the green dashed line illustrates a $<\Delta x^2> \sim t^{}$ relationship. The gray solid line represents the start time of the major merger, while the gray dashed line indicates the end time of the major merger. For Cluster14 and Cluster16, only the start time is marked as the major mergers are still ongoing at redshift zero.}
    \label{fig:tang}
\end{figure*}

We selected particle pairs with separations of 100 kpc $\pm$0.01 kpc 
at \( z = 1.5 \), and performed statistical analysis on the pair dispersion of our four galaxy clusters (Figure~\ref{fig:tang}). We chose a 100 kpc separation for particle pairs to confine our dispersion analysis to select well-resolved pairs within the cluster’s central $\sim$ 500 kpc region while retaining an adequate number of pairs for robust statistics. To examine the possible position dependence of the pair dispersion statistics, we grouped the particle pairs into three radial bins of their initial location:
0-0.5$R_{500\rm{c}}$, 0.5-1$R_{500\rm{c}}$, and 1-2$R_{500\rm{c}}$. 

\begin{figure}
    \includegraphics[width=\columnwidth]{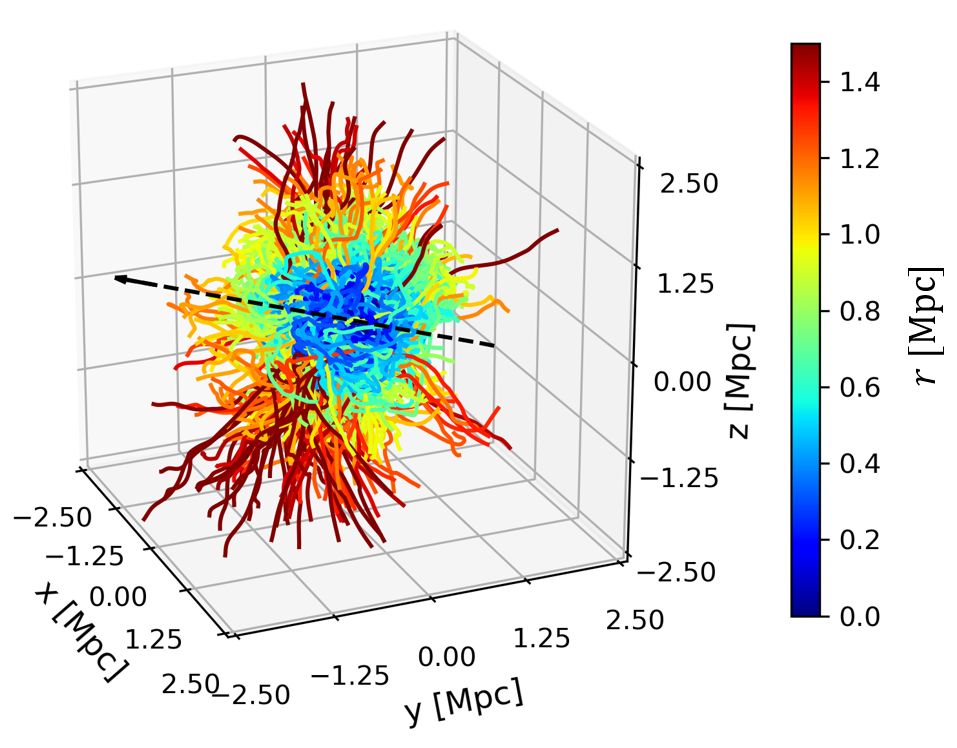}
    \caption{Particle trajectories showing coherent transport during the major merger of Cluster0. The particles are selected at $ a=0.45$ within the range of 0.6-0.8 $R_{500\rm{c}}$. The black dashed line indicates the direction of the major merger. Particles are collectively transported from the inner regions of the galaxy cluster to its outskirts along a specific direction, which is perpendicular to the direction of the major merger.}
    \label{fig:coh}
\end{figure}

\begin{figure}
    \includegraphics[width=\columnwidth]{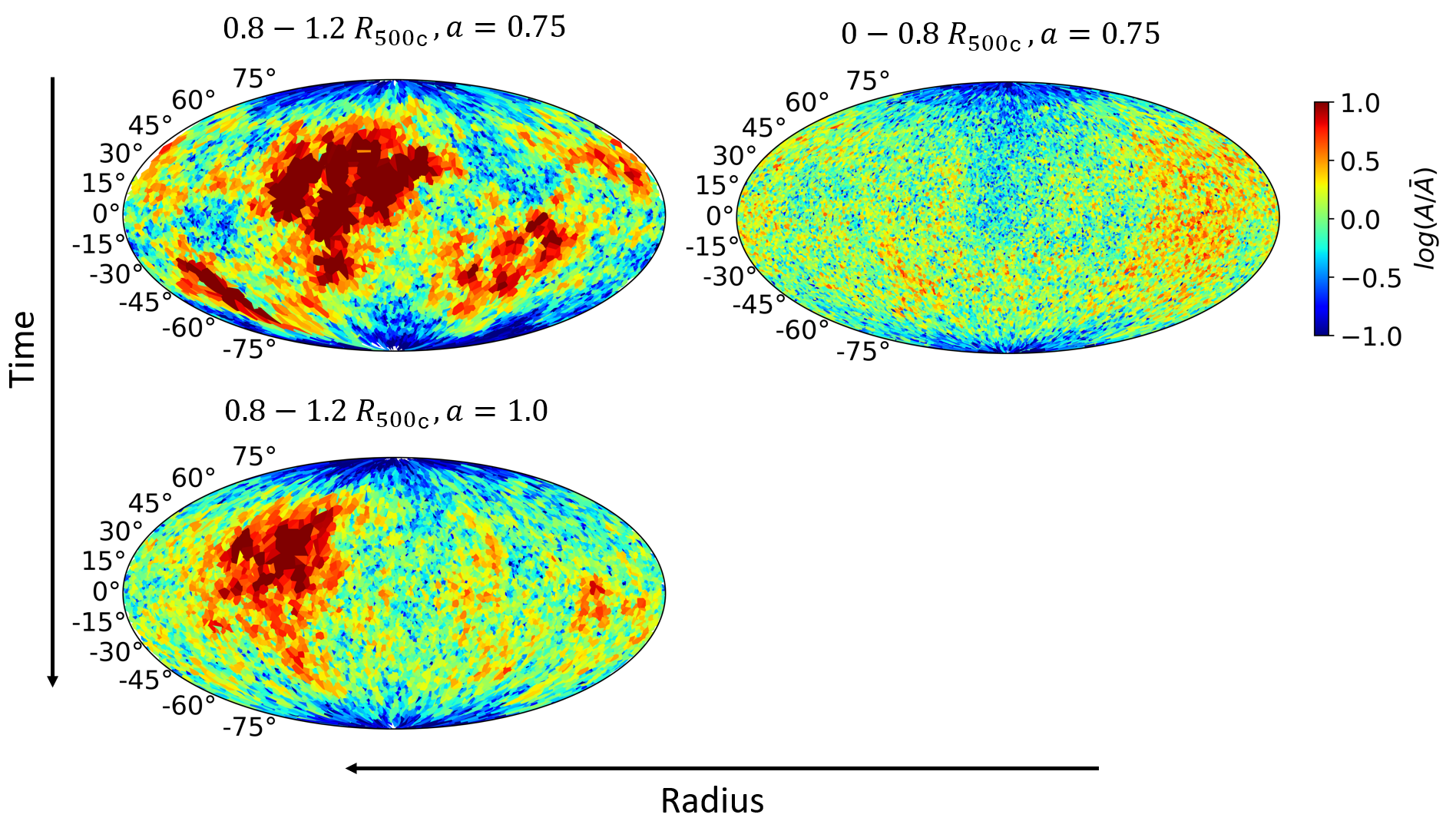}
    \caption{Angular clustering of particles for different bins and different times of Cluster0. The particles are selected within 500 kpc at the start of the major merger. We use the proportion of the area occupied by particles relative to the average area, $A/\bar{A}$, to quantify particle clustering, with the low-density regions associated with larger $A/\bar{A}$ values. Angular clustering shows significant radial dependence and time evolution as a result of the interplay between coherent and turbulent transports.} 
    \label{fig:pos}
\end{figure}

\begin{figure*}
\centering
    \begin{minipage}[b]{1\textwidth}
        \centering
        \includegraphics[width=\linewidth]{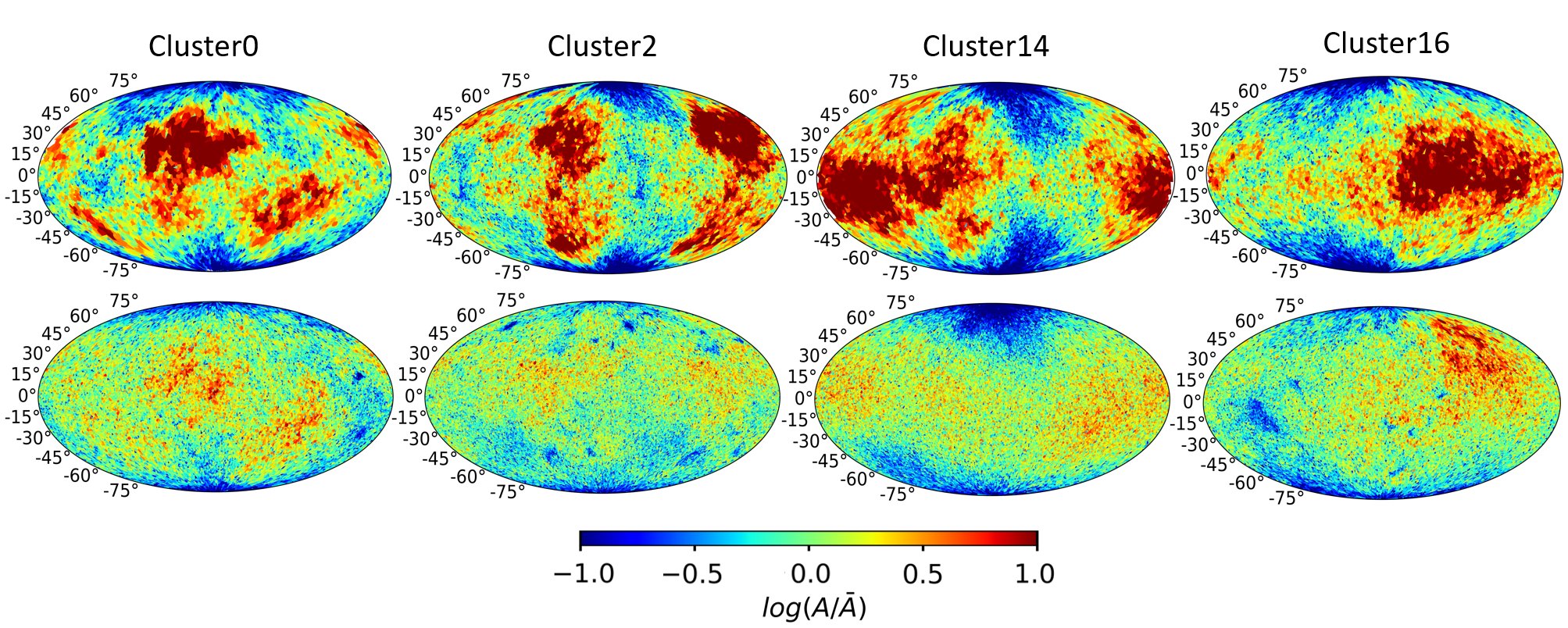}
    \end{minipage}
    \hspace{0.01\textwidth}
    \caption{Angular clustering of particles for within 0.8-1.7 $R_{500\rm{c}}$ at either the end of the major merger (Cluster0, Cluster2) or at the end of the simulation at redshift zero (Cluster14, Cluster16). In the top panels, only the particles within 500 kpc at the beginning of the major merger are shown, while the bottom panels show results for all particles.}
    \label{fig:allbin}
\end{figure*}

\begin{figure}
    \includegraphics[width=\columnwidth]{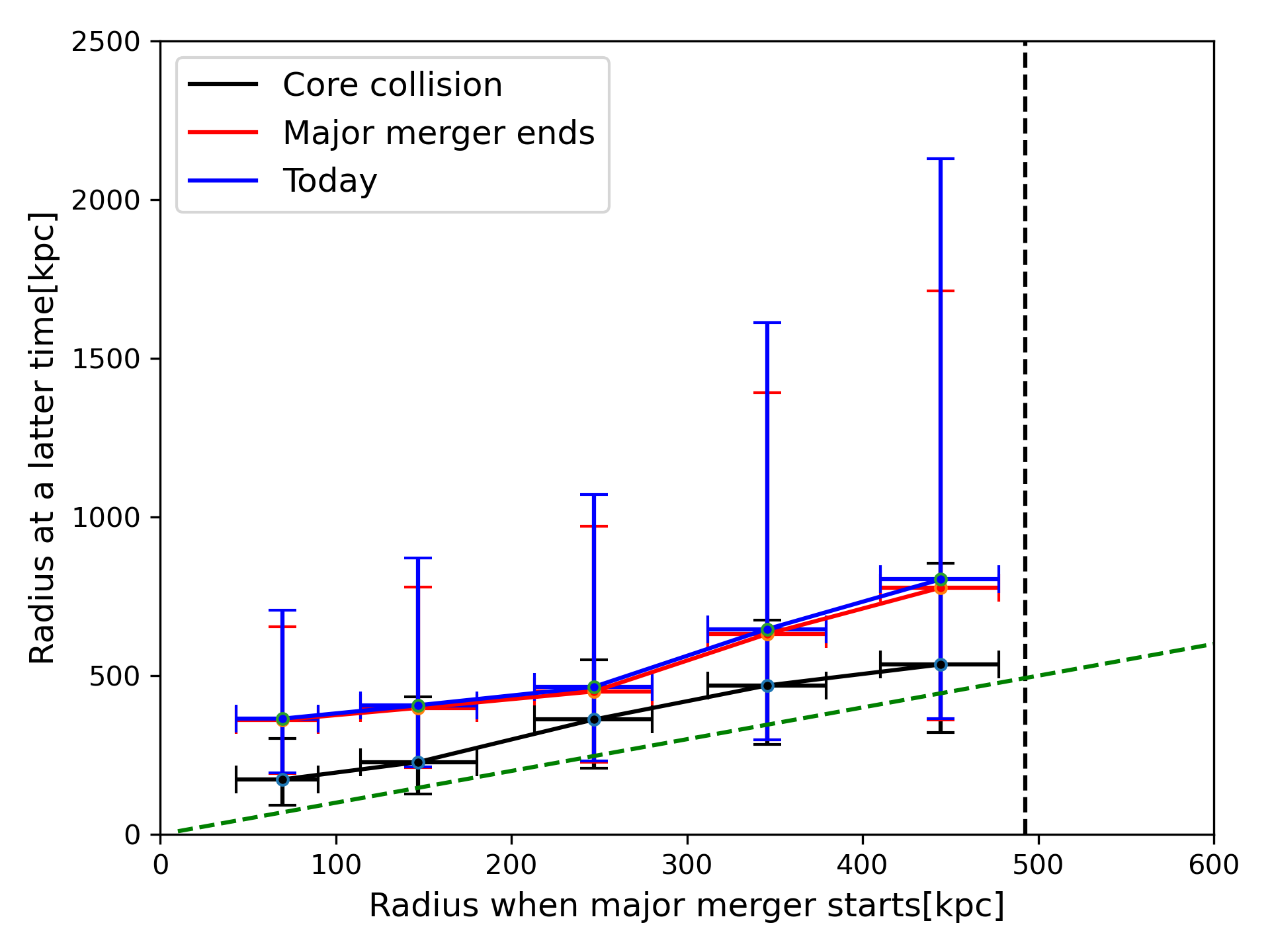}
    \caption{The evolution of the radius of particles for Cluster0. The black dashed line shows the radius of $R_{500\rm{c}}$ at $a=0.45$. The slope of the green dashed line is 1. The central value, lower and upper end of the error bars indicate the 50th,16th and 84th percentiles of the distribution, respectively. During the major merger, the median radius increased greatly. Afterward, when the cluster is relaxed, only the variance but not the median had a significant increase.}
    \label{fig:radius}
\end{figure}

\begin{figure*}
\centering
    \begin{minipage}[b]{1\textwidth}
        \centering
        \includegraphics[width=\linewidth]{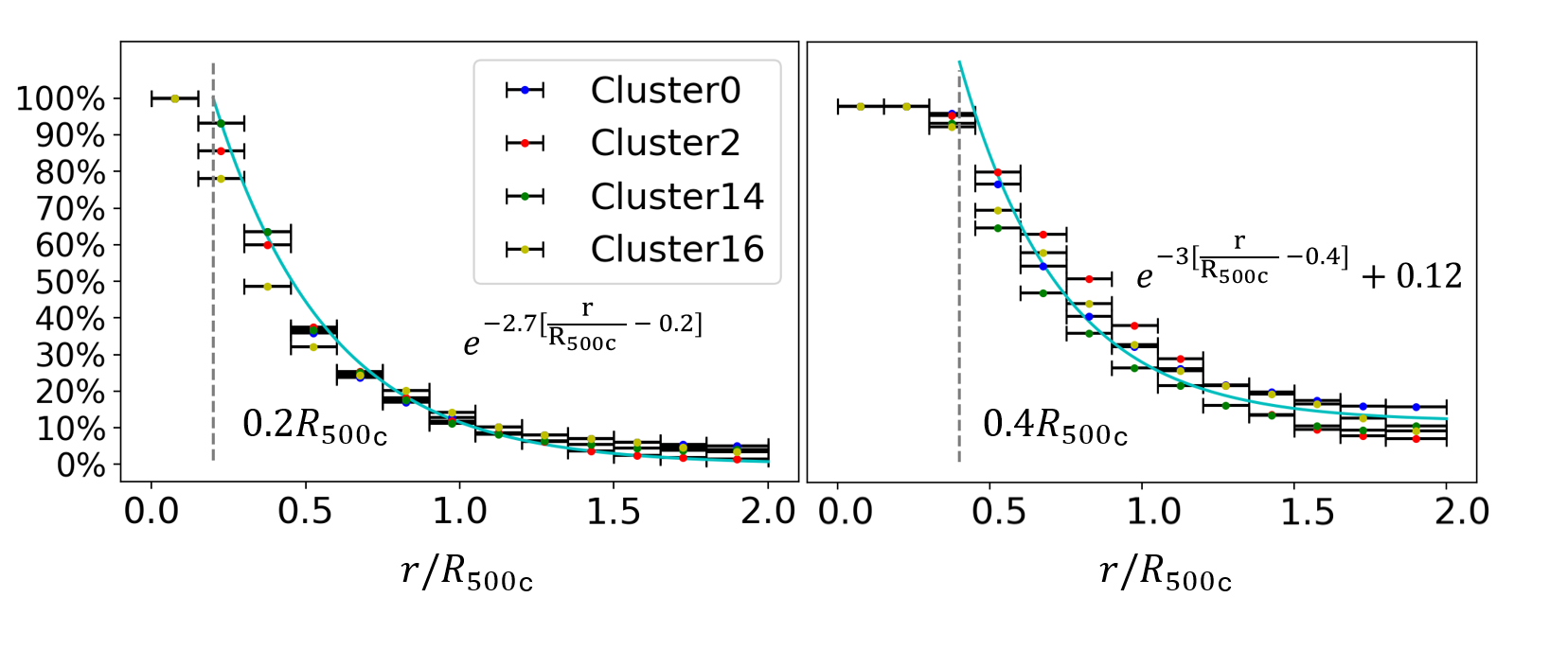}
    \end{minipage}
    \hspace{0.01\textwidth}
    \caption{The fraction of particles that reached the inner region of the galaxy cluster 0.2$R_{500\rm{c}}$ (left) and 0.4$R_{500\rm{c}}$ (right) during the entire evolutionary process. 
    }
    \label{fig:r200kpc}
\end{figure*}

One key result was that all four galaxy clusters showed much faster pair dispersion during the major merger stage than before or after. 
In Figure~\ref{fig:tang}, we marked the beginning of the major merger stage with the solid vertical line and the end with the dashed vertical line. The Richardson scaling was indicated with the red dashed line, and the Brownian diffusion law was indicated as the cyan dashed line. During the major merger stage, the standard deviation of the separation between particle pairs scaled with time with an average slope of approximately 2 for most clusters, i.e., in between those of Richardson and Brownian laws. During the whole major merger stage, one could also identify episodes of rapid increase of the pair separations with slopes comparable, or even greater than the Richardson $\Delta x^2 \propto t^3$ scaling. In the extreme case of Cluster2, a single episode of rapid dispersion resulted in an average slope of $\sim$3 over the entire major merger stage.

Notably, the three radial bins within each cluster showed consistent time behavior but different amplitudes. The change in the slopes of the pair-dispersion statistics was largely synchronous for the three radial bins, manifesting the impact of major events from the outskirts to the central regions of galaxy clusters on a relatively short time scale. On the other hand, the amplitude of $\langle \Delta x^2 \rangle$ in the $0.5{-}1 \, R_{500\rm{c}}$ bin was consistently larger than that of the $0{-}0.5 \, R_{500\rm{c}}$ bin, while $\langle \Delta x^2 \rangle$ in the $1{-}2 \, R_{500\rm{c}}$ bin was generally the largest among all the bins. This was likely a result of faster turbulence delay time scales at smaller radii (\citealt{Shi_2014, Shi_2019}), as is also evident in the evolution of the power spectra of the turbulent ICM motions. In particular, wavelet power spectra of ICM gas motions of Cluster2 and Cluster14 can be found in Figure 4 of \citet{Shi_2018} as those for ClusterA and ClusterB, respectively. The decay of the power spectra amplitudes following their peak at the major merger shows a strong dependence on the radius, consistent with the radial dependence of the pair-dispersion statistics observed here.
To ensure the robustness of our results, we also analyzed the pair dispersion by binning the particles along the tangential direction. The results showed little dependence on the tangential direction.

These results were contrasted with \citet{Vazza_2010} where it was reported that particle pair dispersion adheres consistently to the Richardson scaling throughout their simulations. Instead, our result indicated that the pair-dispersion statistics for the transport process in the ICM did not follow a universal law but rather relied on the specific dynamical states of individual galaxy clusters. Our result also highlighted the importance of major merger events in promoting particle transport in the ICM. Next, we investigate the modes of particle transport during major merger events.

\subsection{Coherent transport\label{subsec:figures}}
To capture the possibly existing coherent transport, we selected particles within a 500~kpc radius at the onset of the major merger and tracked their 3D trajectories during the major merger. 
We divided both time and radial distance into several bins and analyzed the 3D trajectories of particles within each bin for all samples. 

Coherent particle trajectories were evident for particles in all radial bins, demonstrating the existence of a coherent transport mode. As an example, the particle trajectories in the radial bin with the most significant coherent transport for Cluster0 are shown in Figure~\ref{fig:coh}. A significant fraction of particles at the periphery of the galaxy cluster were systematically transported from the central region of the cluster to its outskirts via coherent transport. 
However, we also noticed that the majority of the particles did not reach the periphery of the cluster and their trajectories did not show much directional coherence. These particles may have been disturbed during the transport process, preventing them from reaching the outer regions. Our initial hypothesis attributed this to the strong turbulence within the cluster, which affected particle trajectories.

We quantified the coherent transport mode with the angular clustering of particles. For all particles in a radial bin, we use a Voronoi tessellation of the angular coordinates to determine the area $A$ occupied by each particle and evaluate the ratio of this area to the mean area, $A/\bar{A}$. The angular distribution of $A/\bar{A}$ for Cluster0 was presented in Figure~\ref{fig:pos}. As shown in the top-left panel, the effect of coherent transport was very significant for the radial bin 0.8-1.2 $R_{\rm 500c}$ at the time $a=0.75$ when the major merger ended, and a large fraction of the $4\uppi$ area was occupied by relatively few particles, suggesting a high degree of anisotropy. After the major merger event, the degree of anisotropy weakened with time, as can be seen by comparing the two panels on the left. 
Comparing the two panels at the top that showed the angular distribution of $A/\bar{A}$ for two different radial bins, one could see that the high degree of anisotropy associated with coherent transport existed only for the outer regions, but not for the region well within $R_{\rm 500c}$, as suggested by the particle trajectories in Figure~\ref{fig:coh}.
This loss of anisotropy with time and towards inner regions demonstrated that the coherent transport was transient and discontinuous. Turbulence likely played a significant role in disturbing the transport process in both time and space.

To evaluate the statistical significance of the coherent transport mode, we applied the same methodology to investigate all four galaxy clusters in our sample and compared the angular distribution near $R_{500\rm{c}}$ of particles that had been transported from the central 500 kpc and that of all particles (Figure~\ref{fig:allbin}). 
The selected time was at either the end of the major merger (Cluster0, Cluster2) or at the end of the simulation at redshift zero (Cluster14, Cluster16). The high degree of anisotropy associated with the coherent transport was observed only in the distribution of particles that had been transported from the inner regions, but not in the distribution of all particles. This indicated that, while coherent transport significantly impacted specific regions, its contribution to the overall particle distribution in the galaxy cluster was relatively minor.

\subsection{Statistical analysis of the transport mechanism}
To further characterize the effects of particle transport in the ICM, we conducted two more statistical studies of tracer particles.

First, we analyzed the evolution of the average radii of the particles. The particles were selected within $R_{\rm 500c}$ at the onset of the major merger and grouped into five radial bins. 

The evolution of particle radii for Cluster0 is shown in Figure~\ref{fig:radius}. During the major merger process, 
in particular, between the core collision at $a = 0.55$ and the end of the major merger at $a = 0.75$, particles in all radial bins were transported outward by a significant amount ranging from tens to hundreds of kpc on average. After the major merger concluded, however, the outward transport of particles was much less efficient. Between $a = 0.75$ and the end of the simulation at $a = 1$, the average radii moved outward at most tens of kpc.

However, during this time, the dispersion of the particle radii continued to increase by a large amount. This demonstrates the different roles of coherent and turbulent transport mechanisms: while the average outward motion was largely associated with the coherent transport mode that existed predominantly during the major merger, turbulent transport that continued to function after the major merger likely played a dominant role in increasing the radial dispersion.

Then, we quantified the transport of ICM particles by showing the proportion of particles that ever reached the innermost regions as a function of cluster radius. For all four clusters, we divided the radial range of \(0{-}2R_{500c}\) into 13 bins at \(z = 0\) and evaluated the fraction of particles within each bin that ever entered \(0.2R_{500c}\) and \(0.4R_{500c}\) throughout the entire evolutionary process. The results were generally consistent across all galaxy clusters (Figure~\ref{fig:r200kpc}), with the proportion of particles entering the inner regions decreasing exponentially with radius. For particles near \(R_{500c}\), the fraction of particles that ever entered \(0.2R_{500c}\) and \(0.4R_{500c}\) was 10\%-20\% and 20\%-40\%, respectively. These results could be used to evaluate the effect of central regions on the ICM outskirts.

\section{CONCLUSIONS}\label{sec:conl}

In this paper, we analyze the transport phenomena resulting from the evolution of galaxy clusters due to large-scale structure formation. The hydrodynamic Reynolds number of the ICM, estimated using the classical Braginski viscosity, is not high, and numerical simulations are limited by resolution constraints, leading to relatively low effective Reynolds numbers and small inertial ranges. Given these limitations, we focus primarily on the diffusion effects of fluid motions on relatively large scales. 
Our main findings are as follows.

Galaxy clusters exhibit significantly faster pair dispersion during the major merger stage compared to before or after. The standard deviation of particle pair separations scales with time at an average slope of approximately 2, intermediate between Richardson and Brownian laws, with some episodes exceeding Richardson scaling (\(\Delta x^2 \propto t^3\)). Particles in different radial bins show consistent temporal trends but differ in amplitude, likely due to faster turbulence delay times at smaller radii. Unlike \citet{Vazza_2010}, which reported adherence to Richardson scaling, our findings suggest pair-dispersion in the ICM depends on the specific dynamical state of individual clusters, highlighting the critical role of major mergers in promoting particle transport. 

To investigate the particle transport modes during the major merger, we attempted to capture the potentially existing coherent transport and successfully identified this mechanism. We quantified the coherent transport mode using the angular clustering of particles and found that coherent transport is transient and discontinuous. Turbulence likely caused significant disturbances to the transport process both temporally and spatially. Moreover, we observed that while coherent transport has a notable impact in specific regions, its contribution to the overall particle distribution within the galaxy cluster is relatively minor.

To further characterize the effects of particle transport in the ICM, we performed two statistical studies on the tracer particles. The radial evolution of tracer particles is consistent with the picture that the average outward motion is primarily associated with the coherent transport mode which exists mainly during the major merger process; while after the merger ends, turbulent transport becomes the dominant mechanism in increasing radial dispersion. The analysis of the fraction of particles that ever reached the inner region of the galaxy cluster shows an exponentially decreasing fraction with cluster radius.

All our results consistently highlight the dominant role of major merger events in facilitating particle transport in the ICM, and the interplay between the coherent and turbulent transport mechanisms.

\section{Acknowledgments} \label{sec:cite}

\begin{acknowledgments}

This work is supported by NFSC grant No. 11973036, and YWZ is supported by the Scientific Research and Innovation Project of Postgraduate Students in the Academic Degree of Yunnan University. We thank the anonymous referee for the helpful comments and suggestions.

\end{acknowledgments}

%



\appendix

\section{Resolution Effects on Tracer-Particle Positional Accuracy in ICM Studies}

In our simulation, tracer particles are arranged to sample the density field of the ICM. The finite spatial and temporal resolution of the velocity field that advects them can indeed lead to positional errors, biasing tracer densities relative to the Eulerian gas density. This issue was noted by \citet{Genel2013} and \citet{Cadiou2019}. In low resolution cosmological runs, biases in low mass halos can be as large as an order of magnitude. However, increasing the resolution significantly reduces this bias (see Fig. 13 of \citealt{Genel2013}). Our relatively high resolution in both space and time ensures that this bias is not severe. In a previous work using this tracer particle simulation, we have compared the Eulerian gas density and the tracer particle density (Figs. A1 and A2 of \citealt{Shi_2020}), and found that they are consistent.

\section{Relation between Pair Dispersion and Velocity Structure Function}
The pair dispersion statistic quantifies how the distance between two fluid elements (or tracer particles) evolves over time. This evolution is governed by the relative velocity between the particles, which is characterized by the velocity structure function. For example, the second-order velocity structure function is defined as
\begin{equation}
    S_2(r) = \langle [\mathbf{v}(\mathbf{x} + \mathbf{r}) - \mathbf{v}(\mathbf{x})]^2 \rangle,
\end{equation}
which measures the typical velocity difference at a spatial separation $r$. In the inertial range of turbulence, this structure function typically scales as
\begin{equation}
    S_2(r) \propto r^{2/3}
\end{equation}
following Kolmogorov's theory. This scaling indicates that the relative velocity $\delta v$ between points separated by $r$ behaves as
\begin{equation}
    \delta v \sim r^{1/3}.
\end{equation}

Since the rate of change of the particle separation is directly linked to $\delta v$ (i.e., $dr/dt \sim \delta v$), integrating this relationship leads to the famous Richardson scaling:
\begin{equation}
    r(t) \propto t^{3/2}.
\end{equation}
Thus, the velocity structure function directly informs the growth rate of the pair dispersion statistic, bridging the characteristics of the turbulent velocity field with the observed dispersion behavior.


\bibliography{paper}
\bibliographystyle{aasjournal}



\end{document}